\documentclass[preprint,12pt,3p]{elsarticle}

 \usepackage{graphicx}

\usepackage{amssymb}
\usepackage{amsmath}
\usepackage{caption}
\usepackage{subcaption}
\usepackage{amsfonts}
\usepackage{amssymb}
\usepackage{amsmath}
\usepackage{commath}
\usepackage{mathrsfs}
\usepackage{graphicx}
\usepackage{float}

\journal{}

\begin{document}

\begin{frontmatter}

\title{Transition time estimation for $\delta$-function coupling in two state problem: An analytically solvable model}

\author{Mayank Vashistha, Chinmoy Samanta* and Aniruddha Chakraborty \\ School of Basic Sciences, Indian Institute of Technology Mandi, Kamand, Himachal Pradesh-175075, India}

\begin{abstract}
We propose a simple method to calculate transition time in a two-state scattering problem, where two constant potentials are coupled by a delta function potential $V_{12}=V_{21}=k_0 \delta(x)$. The exact analytical expression for the time of transition $\tau$ is derived. We notice $\tau$ explicitly depends on the second state's potential energy along with the incident energy and coupling strength. We also observe from the derived expression of $\tau$ that depending on the initial energy, the coupling potential could behave like a transparent or opaque medium to the incident wave in a single state equivalent description.   
\end{abstract}


\end{frontmatter}


\section{Introduction}
\label{sec1}
\noindent Multi-state problems are important in many different areas of physics, chemistry or even in biology. The estimation of non-adiabatic transitions probability is reported in $1932$ by Landau \cite{Landau}, Zener \cite{Zener} and Stueckelberg \cite{Stuckelberg} and by Rosen and Zener \cite{Rosen}.  Since then huge number of research papers have been published on this very engrossing topic \cite{Naka, Nikitin, Child, Osherov,NikitinRev, NakaBook,NakaACP, Shaik, Imanishi, Thiel, Yoshimori, Engleman, Mataga}. One of the important problem, where two constant potentials are coupled by an exponential function was investigated by Osherov and Voronin \cite{Osherov2}. The problem where two exponential potentials are coupled by another exponential potential was studied by C. Zhu  \cite{Zhu}.  Earlier publication from our group reported exact analytical solution for those cases where two or more arbitrary potentials are coupled by Dirac delta potential \cite{Chakraborty2009,AC2009,Chakraborty2011,Diwaker}. Even the method can be extended for an arbitrary coupling \cite{Chakraborty2012}. In reality, the transition probability between different potentials is maximum at the crossing point between two potentials. This is because the energy transfer between the electronic and nuclear degrees of freedom is least at the crossing point. Therefore the model where coupling between two potentials is localized in space is more important than a model where the coupling is same everywhere. The most simplest way to represent localized coupling is by using a Dirac Delta function and this delta function coupling model has the advantage that it has an exact analytical solution \cite{Chakraborty2009,AC2009,Chakraborty2011,Diwaker}. \\
In the present study, we consider the simplest possible case, where two constant potentials are coupled by a Dirac delta potential. Using this model we will derive an exact analytical expression  for the transition time, which explicitly posses the effect of second state potential into the first state. It could be realized that in a one dimensional two state system, this quantity is equivalent to the tunneling time in the case of a point interaction in a single state problem of a quantum object. This is where the significance of discussing the tunneling time in our study becomes necessary. It has been observed that due to the presence of second state, there occurs a time delay i.e., the quantum object prefers to spend relatively longer time in the region where coupling is non-zero. Landauer et al., \cite{landauer1989,Buttiker,landauer1994} expressed tunneling time as the gradient of transmission amplitude and phase with respect to potential in a series of papers \cite{landauer1989,Buttiker,landauer1994}. Tunneling time has been a topic of profound debate since many years \cite{Nicolas}. In quantum mechanics there does not exist a time operator to determine time related entities. Non-existence of time operator leads to various names and definitions associated with tunneling time. Over a decade there have been many different theoretical approaches to calculate tunneling time \cite{Nicolas,Yuan,Pollak2017,Pollak2018}. Until recently the debate was purely theoretical, considering the process to be instantaneous for all practical purposes. Development of ultrafast lasers and in particular, the 'attoclock' technique\cite{Keller2019,Landsman,Torlina} addressed this issue and changed the perspective of addressing it theoretically along various experimental approaches which have been claimed and compared with the competing theoretical approaches to gain broad essence on tunneling time \cite{Keller2019,Landsman,Tomas,Moshammer,Satya}. In the following section, we follow the method of tunneling time calculation for a delta potential energy barrier as given by S. Lj. S. Kocinac et al., \cite{Kocinac}. Interestingly there are various different expressions of tunneling time available in literature \cite{kocinac2008} also, among them only dwell time and asymptotic phase time has a unique expression \cite{Kocinac, kocinac2008}. To the best of our knowledge no research work is done so far for estimating the effect of second potential on the transition time into the first potential. 

\section{The Model: Two state problem with $\delta-$function coupling}
\label{sec2}
\begin{figure}[h!]
\centering
\begin{subfigure}{.5\textwidth}
  \centering
  \includegraphics[width=.8\linewidth]{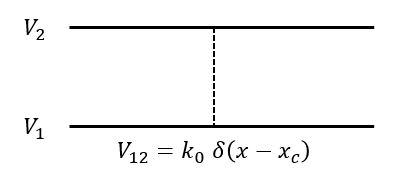}
  \caption{}
  \label{fig:sub1}
\end{subfigure}%
\begin{subfigure}{.5\textwidth}
  \centering
  \includegraphics[width=.8\linewidth]{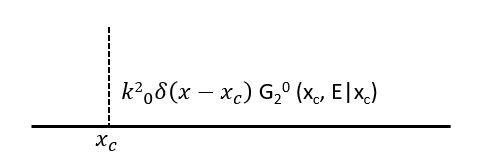}
  \caption{}
  \label{fig:sub2}
\end{subfigure}
\caption{(a) Schematic diagram of two state problem with two constant potentials coupled by a $\delta-$function potential. (b) Schematic diagram of the case  where two state problem is reduced to an effective single state problem with a $\delta-$function potential.}
\label{fig:test}
\end{figure}
\noindent In this section, we consider two state problem involving two constant potentials $V_1$ and $V_2$ and corresponding Hamiltonian as $H_1$ and $H_2$  and the coupling between these two potential is represented by Dirac delta function potential as given below
\begin{equation}
V_{12}(x)= V_{21}(x) = k_0\delta(x-x_c),
\end{equation}
where $k_{0}$ represent the strength of the $\delta-$function potential  and the corresponding time independent Schr\"odinger equation in the matrix form can be written as:
\[
\begin{bmatrix}
    H_{1}       & k_{0}\delta{(x-x_c)}  \\
    k_{0}\delta{(x-x_c)}       & H_{2}  \\

\end{bmatrix} \begin{bmatrix}
    \phi_{1}(x)         \\
    \phi_{2}(x)   \\

\end{bmatrix}
=
 E \begin{bmatrix}
     \phi_{1}(x) \\
    \phi_{2}(x)   \\

\end{bmatrix}.
\]
The above equation in the matrix form can be decomposed into the following two equations
\begin{equation}
H_1\phi_1(x) + k_0\delta{(x-x_c)}\phi_2(x) = E \phi_1(x).
\end{equation}
\begin{equation}
H_2\phi_2(x) + k_0\delta{(x-x_c)}\phi_1(x) = E \phi_2(x).
\end{equation}
Using Eq. (3), $\phi_2(x)$ may be written in the following form
\begin{equation}
\phi_2(x) = k_0[E-H_2 (x)]^{-1}\delta{(x-x_c)}\phi_1(x). 
\end{equation}
Substituting $\phi_2(x)$ in Eq. (2), we get
\begin{equation}
H_1\phi_1(x) + {k_0}^2\delta{(x-x_c)}[E-H_2 (x)]^{-1}\delta{(x-x_c)}\phi_1(x) = E \phi_1(x),
\end{equation}
which further can be represented as
\begin{equation}
H_1\phi_1(x) + {k_0}^2 \delta{(x-x_c)}  \mathscr{G}^{0}_{2}(x_c, x_c|E)\phi_1(x) = E \phi_1(x),
\end{equation}
where $ \mathscr{G}^{0}_{2}(x_1, x_2|E) = \left <x_1|[E-H_2 (x)]^{-1}|x_2 \right> $ is Green's function for the second state of the quantum object with energy $E$ in the absence of coupling. Note that it is the Fourier transformation of $G^{0}_{2}(x_1, x_2|t)$, which is the probability amplitude that a quantum object starting at $x_2$ may be found at $x_1$ at the time $t$. It obeys the following equation
\begin{equation}
\left[ i \hbar\frac{ \partial}{\partial t} - H_{2} \right] G^{0}_{2}(x_1, x_2|t) = \delta (x_1 -x_2).    
\end{equation}
\noindent The above equation does not have any coupling term in it. As there is no coupling term, there is no loss of probability amplitude. Therefore 
\begin{equation}
\int_{- \infty}^{\infty} |G^{0}_{2}(x_2, x_1|t)|^2 dx_2 =1.
\end{equation}
\noindent Now we assume $V_{1}$ to be equal to $0$, so that the Eq. (6) reduces to 
\begin{equation}
- \frac{\hbar^2}{2 m} \frac{\partial^2 \phi_{1}(x)}{\partial x^2} + {k_0}^2\delta{(x-x_c)}  \mathscr{G}^{0}_{2}(x_c, x_c|E)\phi_1(x) = E \phi_1(x).
\end{equation}
In the above, Eq.(9) represents time independent Schr\"odinger equation for $\delta-$function potential  with strength $k_{0}^2  \mathscr{G}^{0}_{2}(x_c, x_c|E)$. Therefore, our two state problem with $\delta-$function coupling is reduced into a single state problem subjected to a simple $\delta-$function potential. The graphical abstract of the situation has been shown in Fig. 1, the left panel describes the old profile and right panel describes the new transformed profile of the two state problem.

\section{Methodology}
\label{sec3}
\noindent 
 We have started with previously existed method of calculating tunneling time in the single state scattering problems for our aimed estimation of transition time in two sate problem. The relation among relevant tunneling times is stated as \cite{Kocinac, kocinac2008}
\begin{equation}
\tau_d = \tau_a +\tau_g - \tau_i ,
\end{equation}
 where $\tau_d$ is dwell time, which represents total spending time of the particle in the potential barrier. The second term $\tau_a$ in Eq. (10) describes as a absorption term depending explicitly on the imaginary part of the potential. The following term, the bidirectional group delay time $\tau_g$ could be calculated from 
\begin{equation}
\begin{split}
    \tau_g = {\abs{T(E)}}^2 \tau_{gt} + {\abs{R(E)}}^2 \tau_{gr},\\
   \tau_{gt}=\hbar \frac{d\phi_t}{dE}\quad \text{and}\quad \tau_{gr}=\hbar \frac{d\phi_r}{dE}.
\end{split}
\end{equation}
In the above, $T(E)$ and $R(E)$ are transmission and reflection coefficients respectively as a function of energy. $\phi_t$ and $\phi_r$ are being phases of $T(E)$ and $R(E)$ respectively. The last term, self interaction time $\tau_i$ appears due to the overlap of the incident and reflected part of the wave in front of the barrier. It could be shown that for the $\delta-$function potential, 
\begin{equation}
\tau_d = \tau_a = 0
\end{equation}
and we remain with $\tau_g = \tau_i$. We consider the coupling location to be at the origin ($x_c = 0$), so the time independent Schr\"odinger equation is given by
\begin{equation}
- \frac{\hbar^2}{2 m} \frac{\partial^2 \phi_{1}(x)}{\partial x^2} + {k_0}^2\delta{(x)}  \mathscr{G}^{0}_{2}(0, 0|E)\phi_1(x) = E \phi_1(x).
\end{equation}
\noindent The equation in the region $x \ne 0$ is reduced to
\begin{equation}
\frac{\partial^2 \phi_{1}(x)}{\partial x^2} = - k^2 \phi_1(x),
\end{equation}
with $ k^2 =2m E/\hbar^2 $. For the left side of the potential  barrier, the solution of the above equation is given by
\begin{equation}
\phi_1(x) = e^{ i k x} + B e^{- i k x},
\end{equation}
\noindent and on the right side of the potential the solution is given by
\begin{equation}
\phi_1(x) = C e^{ i k x}.
\end{equation}
We can now find these two unknowns $B$ and $C$ by using the following two boundary conditions
\begin{equation}
\phi_1(0+\epsilon)=\phi_1(0- \epsilon),    
\end{equation}
\noindent and
\begin{equation}
\frac{ \partial \phi_1(x)}{\partial x}|_{x= 0+\epsilon} - \frac{ \partial \phi_1(x)}{\partial x}|_{x = 0 - \epsilon} =\frac{2 m}{\hbar^2} k_0^2  \mathscr{G}^{0}_{2}(0, 0|E)\phi_1(0).    
\end{equation}
\noindent Using Eq. (17) and Eq. (18) one can easily calculate the coefficients $B$ and $C$. Hence the explicit form of transmission and reflection coefficients and associated phases are given by
\begin{equation}
\begin{split}
    |T(E)|^2 = \frac{\hbar^4 k^2}{\left(\hbar^2 k \right)^2 + \left(m {k_0}^2 \mathscr{G}^{0}_{2}(0, 0|E)\right)^2},\quad & |R(E)|^2 = \frac{\left(m {k_0}^2 \mathscr{G}^{0}_{2}(0, 0|E)\right)^2}{\left(\hbar^2 k \right)^2 + \left(m {k_0}^2 \mathscr{G}^{0}_{2}(0, 0|E)\right)^2}\\
    \phi_t=\tan^{-1}\left( - \frac{m {k_0}^2\mathscr{G}^{0}_{2}(0, 0|E)}{\hbar^2k }\right),\quad& \phi_r=\tan^{-1}\left(  \frac{\hbar^2k }{m {k_0}^2\mathscr{G}^{0}_{2}(0, 0|E)}\right).
\end{split}
\end{equation}
As mentioned in Eq. (12), dwell time and absorption time are zero for $\delta-$function potential, only group delay time $\tau_g$ will be survived. The Greens function $\mathscr{G}^{0}_{2}(x, x_{c}|E)$ could be found by solving the equation \cite{Grosche}
\begin{equation}
    [E-H_{2}(x)]\mathscr{G}^{0}_{2}(x, x_{c}|E)=\delta(x-x_C)
\end{equation}
and for the constant second state potential $V_{2}=V$ ($>E$), it is given by
\begin{equation}
 \mathscr{G}^{0}_{2}(x, x_{c}|E) =-\sqrt{\frac{m}{2 \hbar^2}}\frac{e^{-\sqrt{2m(V-E)/\hbar^2}|x-x_{c}|}}{\sqrt{V-E}}.
\end{equation}
Now our main studying equation in Eq. (13) becomes
\begin{equation}
- \frac{\hbar^2}{2 m} \frac{\partial^2 \phi_{1}(x)}{\partial x^2} - \alpha \delta{(x)}\phi_1(x) = E \phi_1(x),
\end{equation}
where $\alpha= - {k_0}^2\mathscr{G}^{0}_{2}(0, 0|E)$ is a positive function of energy and coupling strength. Here $\alpha$ represents the strength of the $\delta-$function potential well and incident wave is scattered by this potential. Now the problem is reduced to determination of transmission time of the incident wave across the $\delta-$ potential well placed at the origin in a single state problem. In the following, we proceed to calculate the transition time.
As per Eq. (11), $\tau_g$ depends on $T(E)$ and $R(E)$ along with $\tau_{gr}$ and  $\tau_{gt}$, whereas last one is calculated and it is given by 
\begin{equation}
\tau_{gt} = \frac{m \hbar^3 k_0^2}{\sqrt{E}\sqrt{V-E}}\frac{2 E-V}{\left(4 \hbar^4 E(V-E) +k_0^4 m^2\right)}.
\end{equation}
As the potential under consideration is real and symmetric, we find that $\tau_{gt} = \tau_{gr}=\tau_{g}$. For a one dimensional point interaction i.e., a $\delta-$function potential, tunneling time is defined as the obtained group delay in Eq. (20). Since in our case the model consist of two state with a $\delta-$function coupling and the system is closed, the transition time from the ground state to excited state is considered to be the same tunneling time in a one dimensional point interaction. So the transition time is expressed as
\begin{equation}
\tau = \frac{m \hbar^3 k_0^2}{\sqrt{E}\sqrt{V-E}}\frac{2 E-V}{\left(4 \hbar^4 E(V-E) +k_0^4 m^2\right)}.
\end{equation}
With choosing $\hbar=1$, $2m=1$, and $\epsilon=E/V$, the expression for the transition time becomes
\begin{equation}
\tau = \frac{2  (2 \epsilon -1)}{\sqrt{\epsilon}\sqrt{1-\epsilon }(k_{0}^2+16\epsilon(1-\epsilon)(\frac{V}{k_{0}})^2)},
\end{equation}
and the expressions of probabilities in Eq. (19) get the following form
\begin{equation}
|T(\epsilon)|^2 = \frac{1}{1+\frac{k_0^4}{16V^2\epsilon(1-\epsilon)}} ~ ,~ |R(\epsilon)|^2 =  \frac{1}{1+\frac{16V^2\epsilon(1-\epsilon)}{k_0^4}}. 
\end{equation}
\section{Results and Discussion}

\begin{figure}
\centering
\includegraphics[width=.5\linewidth]{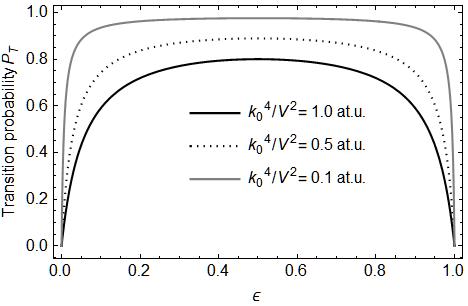}
\caption{Transition probability $P_T=|T(\epsilon)|^2$ against the energy $\epsilon$ for three different values of the parameter $k_0^4/V^2$ (at. u.).}
\label{fig:test3}
\end{figure}
\label{sec4}
 \noindent In this section, we have analyzed the dependency of transition time for simple constant second state potential energy on the incident energy as well as on the coupling strength. Fig. 3(a) shows the variation of transition time with respect to incident energy. As $\epsilon \rightarrow 0$, $\tau$ approaches to negative infinity, and then the coupling potential is near to being opaque. This negative time values appear due to the lagging phase of the transmitted wave with increasing incident energy. When $\epsilon \rightarrow 1$, then $\tau$ reaches to positive infinity, suggesting the coupling potential is almost being opaque. On the other hand, the potential is being transparent when the value of $\epsilon$ lies in the mid-region of two extreme values of it. This nature of the transition time can be understood clearly from the probabilistic profile in Fig. 2, where we only get the maximum transition probability around $\epsilon=0.5$ and zero probability at its two extreme values, which is due to the opaqueness of the coupling potential. In the same figure, we also see that the probability reaches to $1$ around $\epsilon=0.5$ when $V>>\frac{k_{0}^{2}}{4}$. Fig. 3(b) simply describes the variation of transition phase $\phi_t$ with the energy $\epsilon$. The variation looks like the mirror image of that of the probabilistic characteristic in Fig. 2. We already know that the transition time is proportional to the slope of the curve in Fig. 3(b), thus the sign of the transition time is solely determined by the sign of the slope. Pointedly the slope of the curve is positive when $\epsilon \geq 0.5$ and in this region the phase is leading with the energy, giving a positive value of $\tau$. The slope is negative when $\epsilon \leq 0.5$ and here the phase is lagging with the energy, giving a negative value of $\tau$, which is shown in Fig. 3(a). Fig. 4, the plot of $\tau$ vs. $k_{0}^{2}$ describes how the transition time depends on the strength of the coupling potential. We observed that the smaller values of the strength $k_{0}$ ($k_{0}^2< 4V \sqrt{\epsilon(1-\epsilon)}$) (in this figure the range where $k_{0}^2< 2$) hinder to decrease the transition time. The transition time takes a maximum value at $k_{0}^2= 4V \sqrt{\epsilon(1-\epsilon)}$. But higher values of $k_{0}$ ($k_{0}^2> 4V \sqrt{\epsilon(1-\epsilon)}$) help to decrease the transition time. In this figure, it is also seen that the influence of the incident energy is getting less when its value approaches the potential energy ($E\rightarrow V$). The explained characteristics can easily be understood from the expression in Eq. (24) itself. 

\begin{figure}[hbt!]
    \begin{subfigure}{.5\textwidth}
    \centering
   \includegraphics[width=.9\linewidth]{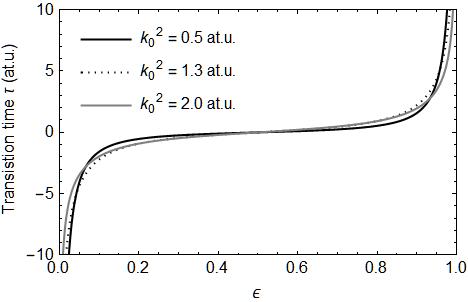}
    \caption{}
\end{subfigure}    
\begin{subfigure}{.45\textwidth}
    \centering \includegraphics[width=.9\linewidth]{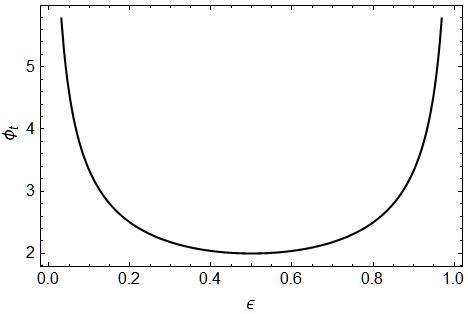}
    \caption{ }
\end{subfigure}  
    \caption{(a) Transition time $\tau$ against incident energy $\epsilon$ for different values of the square of the strength $k_0^2$. The value of the potential $V$ is set to 1 (at.u.). (b) Plot of phase $\phi_t$ versus energy $\epsilon$ with $k^{2}_{0}/4V=1$(at.u.).}
\end{figure}

\begin{figure}[hbt!]
\centering
\includegraphics[width=.5\linewidth]{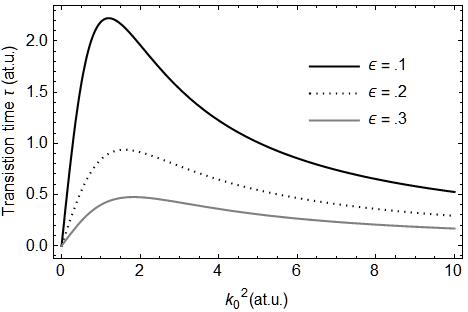}
\caption{Transition time $\tau$ against the square of the coupling strength $k_{0}^{2}$ for four different values of the incident energy $\epsilon$. Here both $\tau$ and $k_{0}^{2}$ (dimension is inverse of time) are plotted in atomic unit (at.u.). The value of the potential $V$ is set to 1  (at.u.). }
\label{fig:test2}
\end{figure}

\section{Conclusions}
\label{sec5}
\noindent We have derived an exact analytical expression of the transition time for the case where two constant potential surfaces are coupled by a $\delta-$function potential in a two-state system. The leading equation in Eq. (9) shows that the effect of second state potential energy is incorporated through the Green's function of the same and thus the transition time does depend on the second state as well. Mainly we have estimated the transition time based on the calculation of group delay time due to the presence of potential barrier or well. For the described model of this two-state system, we conclude the following points regarding this time calculations: \\ 
First, the coupling potential is nearly being opaque in two situations, when energy $E=0$ and $E=V$ ($\epsilon=1$) in an equivalent single state description in Eq. (22). As a result, the transition time $\tau$ approaches to infinity. But the coupling potential is nearly being transparent around the energy $E=V/2$ ($\epsilon=0.5$). This dual nature of the encountered potential has also been observed before in tunneling of an electron through a delta potential barrier \cite{er2005electron}.\\
Second, $\tau$ takes a maximum value when the square of coupling strength $k_{0}^2= 4V \sqrt{\epsilon(1-\epsilon)}$. $\tau=0$ as $k_0^2\rightarrow0$ or $k_0^2\rightarrow\infty$ from Eq. (25).\\
The proposed method is also quite general and can be generalized to the case, where two constant potential surfaces are coupled by a function of arbitrary shape so that the function could be composed by a collection of Dirac delta function with a proper choice of individual strengths \cite{Mayank}.

\section{Acknowledgements}
\noindent The authors acknowledge the research environment and facilities provided by Indian Institute of Technology Mandi. One of the author C.S. would like to thank the institute for providing with Half Time Research Assistantship (HTRA) fellowship.   





\begin{thebibliography}{9}
\bibitem{Landau} L. D. Landau, Phys. Zts. Sowjet., {\bf 2} (1932) {\it 46}.


\bibitem{Zener} C. Zener, Proc. Roy. Soc. A, {\bf 137} (1932) {\it 696}.


\bibitem{Stuckelberg} E. C. G. Stuckelberg, Helv. Phys. Acta, {\bf 5}, (1932) {\it 369}.


\bibitem{Rosen} N. Rosen and C. Zener, Phys. Rev. {\bf 40} (1932) {\it 502}.
\bibitem{Naka} H. Nakamura, Int. Rev. Phys. Chem. {\bf 10} (1991) {\it 123}.

\bibitem{Nikitin} E. E. Nikitin and S. Y. Umanskii,  \textit{Theory of Slow Atomic Collisions} (Springer, Berlin, 1984).

\bibitem{Child} M. S. Child, \textit{Molecular Collision Theory} (Dover, Mineola, New York, 1996) .

\bibitem{Osherov} E. S. Medvedev and V. I. Osherov, \textit{Radiationless Transitions in Polyatomic Molecules} (Springer, New York, 1994).


\bibitem{NikitinRev} E. E. Nikitin, Annu. Rev. Phys. Chem. {\bf 50} (1999) {\it 1}.

\bibitem{NakaBook} H. Nakamura, \textit{Nonadiabatic Transition: Concepts, Basic Theories and Applications} (World Scientific, Singapore, 2002).

\bibitem{NakaACP} H. Nakamura, in \textit{Theory}, \textit{Advances in Chemical Physics}, edited by M. Bayer and C. Y. Ng (John Wiley and Sons, New York, 1992).

\bibitem{Shaik} S. S. Shaik and P. C. Hiberty, edited by Z. B. Maksic, in
\textit{Theoretical Models of Chemical Bonding}, Part 4, (Springer-Verlag, Berlin, 1991), Vol. 82.

\bibitem{Imanishi} B. Imanishi and W. von Oertzen, Phys. Rep. {\bf 155} (1987) {\it 29}.

\bibitem{Thiel} A. Thiel, J. Phys. G {\bf 16} (1990) {\it 867}.

\bibitem{Yoshimori} A. Yoshimori and M. Tsukada, in \textit{Dynamic Processes and Ordering on Solid Surfaces}, edited by A. Yoshimori and M. Tsukada (Springer-Verlag, Berlin, 1985).

\bibitem{Engleman} R. Engleman, \textit{Non-Radiative Decay of Ions and Molecules in Solids}
(North-Holland, Amsterdam, 1979).

\bibitem{Mataga} N. Mataga, in \textit{Electron Transfer in Inorganic, Organic and Biological Systems, Advances in Chemistry}, edited by J. R. Bolton, N. Mataga and G. Mclendon (American Chemical Society, Washington DC, 1991).



\bibitem{Osherov2} V. I. Osherov and A. I. Veronin, Phys. Rev. A, {\bf 49} (1994) {\it 265}.

\bibitem{Zhu} C. Zhu, J. Phys. A, {\bf 29} (1996) {\it 1293}.


\bibitem{Chakraborty2009} A. Chakraborty, Mol. Phys., {\bf 107} (2009) {\it 165}.


\bibitem{AC2009} A. Chakraborty, Mol. Phys., {\bf 107} (2009) {\it 2459}.


\bibitem{Chakraborty2011} A. Chakraborty, Mol. Phys., {\bf 109} (2011) {\it 429}.

\bibitem{Diwaker} Diwaker and A. Chakraborty, Mol. Phys., {\bf 110} (2012) {\it 2257}.

\bibitem{Chakraborty2012} Diwaker and A. Chakraborty, Mol. Phys., {\bf 110} (2012) {\it 2197}.

\bibitem{landauer1989} R. Landauer, Nature, {\bf 341} (1989) {\it 567}.

\bibitem{Buttiker} M. Buttiker, \textit{Electronic Properties of Multilayers and Low Dimensional Semiconductor Structures}, edited by J. M. Chamberlin, L. Evas, J. C. Portal (Springer, New York, 1999), {\it 297}

\bibitem{landauer1994} R. Landauer and T. Martin, Mod. Rev. Phys. {\bf 66} (1994) {\it 217}.

\bibitem{Nicolas} N. Teeny, C. H. Keitel and H. Bauke, Phys. Rev. A, {\bf 94}, (2016) {\it 022104}.
\bibitem{Yuan} M. Yuan, P. Xin, T.  Chu and H. Liu, Opt. Express, {\bf 25} (2017) {\it 19}.
\bibitem{Pollak2017} E. Pollak, J. Phys. Chem. Lett., {\bf 8} (2017) {\it 352}. 
\bibitem{Pollak2018} J. Petersen, E. Pollak, J. Phys. Chem. A,  {\bf 122} (2018) {\it 3563}.
\bibitem{Keller2019} C. Hofmann, A. S. Landsman, U. Keller, J. Mod. Opt, {\bf 66} (2019) {\it 1052-1070}.
\bibitem{Landsman} A. S. Landsman, M. Weger, J. Maurer, R. Boge, A. Ludwig, S. Heuser, C. Cirelli, L. Gallmann, U. Keller,  Optica, {\bf 1} (2014) {\it 343-349}. 
\bibitem{Torlina} L. Torlina  {\it et. al.}, Nat. Phys. {\bf 11} (2015) {\it 503}.
\bibitem{Tomas} T. Zimmermann, S. Mishra, B. R. Doran, D. F. Gordon, A. S. Landsman, Phys. Rev. Lett., {\bf 116}, (2016) {\it 233603}.
\bibitem{Moshammer} N. Camus, E. Yakaboylu, L. Fechner, M. Klaiber, M. Laux, Y. Mi, K. Z. Hatsagortsyan, T. Pfeifer, C. H. Keitel, R. Moshammer, Phys. Rev. Lett. {\bf 119} (2017) {\it 023201}.

\bibitem{Satya} U. S. Sainadh, H. Xu, X. Wang, A-T-Noor, W. C. Wallace, N. Douguet, A. W. Bray, I. Ivanov, K. Bartschat, A. Kheifets, R. T. Sang,  I. V. Litvinyuk {https://arxiv.org/abs/1707.05445} (2017).
\bibitem{Kocinac} S. Lj. S. Kocinac and V. Milanovic, Mod. Phys. Lett. B, {\bf 26} (2012).
\bibitem{kocinac2008} S. Lj. S. Kocinac and V. Milanovic, Phys. Lett. A, {\bf 372} (2008) {\it 191}.

\bibitem{Grosche} C. Grosche and F. Steiner, Springer Tracts Modern Phys. {\bf 145} (1998) {\it 174}.
\bibitem{er2005electron}
B. Er-Juan and S. Qi-Qing, Chin. Phys., {\bf 14} (2005) {\it 208}.


\bibitem{Mayank} M. Vashistha and A. Chakraborty (unpublished).






 












\end{thebibliography}

\end{document}